# Temperature evolution of the effective magnetic anisotropy in the MnCr$_2$O$_4$ spinel

**Dina Tobia[1,2], Julián Milano[1,2], María Teresa Causa[1] and Elin L. Winkler[1,2]**

[1] Centro Atómico Bariloche, CNEA, 8400 S.C. de Bariloche, Río Negro, Argentina

[2] Consejo Nacional de Investigaciones Científicas y Técnicas (CONICET), Argentina

**Abstract**

In this work we present a study of the low temperature magnetic phases of polycrystalline MnCr$_2$O$_4$ spinel through dc magnetization and ferromagnetic resonance spectroscopy (FMR). Through these experiments we determined the main characteristic temperatures: $T_C \sim 41$ K and $T_H \sim 18$ K corresponding, respectively, to the ferrimagnetic order and to the low temperature helicoidal transitions. The temperature evolution of the system is described by a phenomenological approach that considers the different terms that contribute to the free energy density. Below the Curie temperature the FMR spectra were modeled by a cubic magnetocrystalline anisotropy to the second order, with $K_1$ and $K_2$ anisotropy constants that define the easy magnetization axis along the <110> direction. At lower temperatures, the formation of a helicoidal phase was considered by including uniaxial anisotropy axis along the [1$\bar{1}$0] propagation direction of the spiral arrange, with a $K_u$ anisotropy constant. The values obtained from the fittings at 5 K are $K_1 = -2.3 \times 10^4$ erg/cm$^3$, $K_2 = 6.4 \times 10^4$ erg/cm$^3$ and $K_u = 7.5 \times 10^4$ erg/cm$^3$.

**1. Introduction**

The cubic spinels $AB_2O_4$, where the tetrahedral A-sites are occupied by non-magnetic ions and the octahedral B-sites are occupied by Cr ions, are model systems to study magnetic frustration [1, 2, 3]. In these compounds the main magnetic interaction is the strong $J_{CrCr}$ antiferromagnetic direct exchange between the nearest neighbors ions [4, 5]. However, the geometrical arrangement of these magnetic ions in a pyrochlore-like array prevents the magnetic order till very low temperature, as compared to the Curie temperature, $\Theta_{CW}$ [2, 6, 7]. Several authors have proposed that trough the magnetoelastic coupling the strong magnetic frustration could be released and the system could develop a magnetic transition [8,9]; in fact the low temperature magnetic ordered state is usually accompanied by structural distortions. [10, 11] Instead, when the tetrahedral A-site is occupied by a magnetic ion, the magnetic frustration is partially relieved by the $J_{ACr}$ superexchange interaction. [12] In this case the system presents nearly degenerated ground states and it develops complex low temperature magnetic order.

In particular in the $MnCr_2O_4$ the competing Cr-Cr, Cr-Mn and Mn-Mn exchange interactions prevent the development of ferrimagnetic order till to $T_C$~41 K, even considering the important exchange energies observed ($\Theta_{CW}/T_C>10$) [4]. Neutron diffraction studies reported that below $T_C$ the system presents long-range ferrimagnetic order with an easy axis parallel to the <110> direction [13-15], when the temperature decreases below $T_H$~18 K, this magnetic phase coexists with short-range spiral order. In the spiral arrange two positions can be distinguished for the Cr, and the magnetic moments describe a cone on each sublattice, with helicoidal propagation vector in the $[1\bar{1}0]$ direction. The complex low temperature order, where the spin rotation axis does not coincide with the helicoidal propagation vector, positioned this material as a good candidate to present magnetodielectric coupling [16-18]. Recently, Mufti and collaborators [19, 20] have reported that the dielectric and magnetic properties are coupled below $T_H$ in powder MnCr$_2$O$_4$ oxide. In addition, recent FMR results on frustrated spinels [21] have related the unusual FMR temperature dependence to phase separation. In this complex scenario the ferromagnetic resonance (FMR) spectroscopy emerges as a suitable technique because it provides microscopic information related to the exchange and magnetic anisotropy and allows extending the knowledge of the nature of the long-range ferrimagnetic order and the spiral short-range state. In this context we present a study of the low temperature magnetic phases in a cubic chromium spinel with A=Mn by magnetic and FMR measurements. We follow the temperature evolution of the parameters that characterize the FMR spectra in a polycrystalline sample. We describe the evolution of the FMR spectra by a phenomenological model that takes into account the different terms that contribute to the magnetic anisotropy of the system.

## 2. Experimental

Single phase polycrystalline samples of MnCr$_2$O$_4$ were fabricated by solid state reaction of MnO and Cr$_2$O$_3$ powders, as described elsewhere [4]. This system has a normal cubic spinel structure, belonging to the Fd-3*m* space group. The magnetic properties were investigated on loosely packed powdered samples in the 5–90 K temperature range, with applied fields up to 5 T, using a commercial superconducting quantum interference device (SQUID, Quantum Design MPMS-5S) magnetometer. The temperature dependence of the ferromagnetic resonance (FMR) spectra was recorded by a Bruker ESP300 spectrometer operating in the conventional absorption mode at $\omega/2\pi \sim 24$ GHz (K-band), for temperatures ranging from 4 K to 300 K. Magnetic-field scans were performed in the range 0 – 15000 Oe. Care was taken in order to avoid cavity detuning effects, as are usually present in spectra of strongly magnetic compounds. For that purpose, the $MnCr_2O_4$ powder was thoroughly milled and mixed with a non-absorbing KCl salt. No noticeable changes in the quality factor (Q) of the cavity were registered in the whole set of experiments.

## 3. Results and discussion

*3.1 Magnetic properties*

Figure 1 presents the magnetization vs. temperature measurements, M(T), under zero-field-cooling (ZFC) and field-cooling (FC) conditions, with an applied field of 50 Oe. Near 41 K, a sudden jump is observed, consistent with the ferrimagnetic transition ($T_C$). As the temperature is further lowered, other anomalies are manifested at $T_H$~18 K and $T_f$~14 K, corresponding, respectively, to the helicoidal order temperature and to the "lock-in" transition at which the spiral becomes fully developed, as it was determined from neutron diffraction experiments [13-15]. The inset in figure 1 exhibits the M(T) ZFC-FC curves measured with an applied field of 8 kOe, where it can be observed that the $T_C$ value increases and the transition becomes broader. Also, when the applied magnetic field is enhanced, both low-temperature anomalies become less defined, as it was previously reported by Mufti et al. [19, 20].

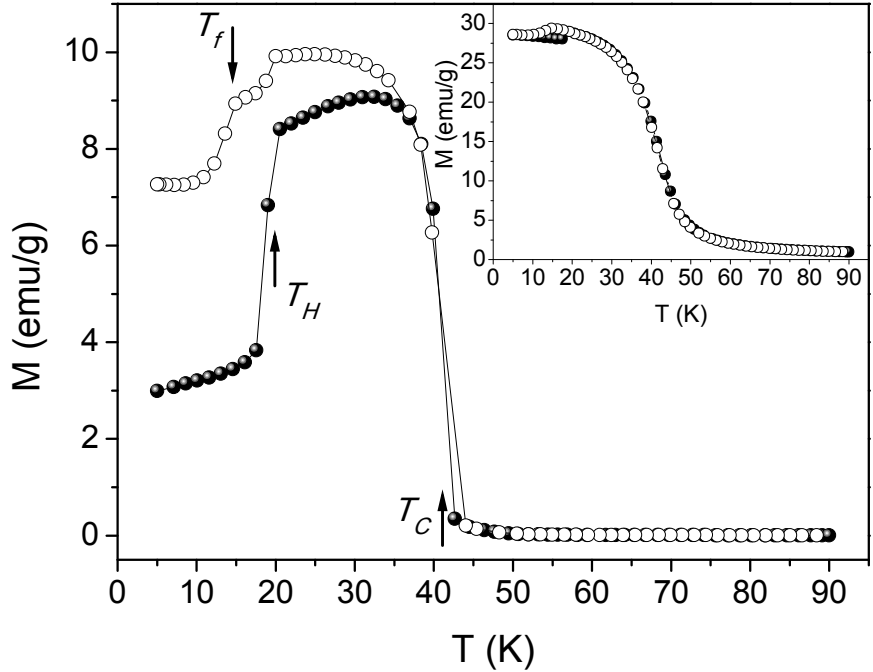

**Figure 1.** Temperature dependence of the ZFC (solid symbols) and FC (open symbols) magnetization measured in a field of 50 Oe. The arrows signal the ferrimagnetic transition ($T_C$), the helicoidal order temperature ($T_H$) and the "lock-in" transition where the spiral component is fully developed ($T_f$). The inset shows the M(T) ZFC-FC curves measured with an applied field of 8 kOe.

Figure 2 shows the magnetization as a function of the applied magnetic field acquired at different temperatures. As the temperature descends below ~45 K the magnetization presents an important increase that starts near 2.5 kOe. The spontaneous magnetization of $MnCr_2O_4$ at 5 K was estimated to be ~1.1 $\mu_B$ per unit formula in agreement with the value previously reported [19,20,22]. Noticeable, a linear increase of the high field magnetization is clearly observed for temperatures below 30 K. This lineal contribution signals a non-collinear spins arrangement of the $MnCr_2O_4$ ferrimagnet. As is stated in references [23, 24] in non-collinear configuration the applied magnetic field exerts a torque that could change the angles between the canted magnetic moments; as a result the magnetization increases linearly with the magnetic field. By neutron diffraction studies non-collinear order was found below T~18 K where short-range spiral arrangement is developed [13,14]. In order to shed light onto this complex behavior we have performed ferromagnetic resonance measurements.

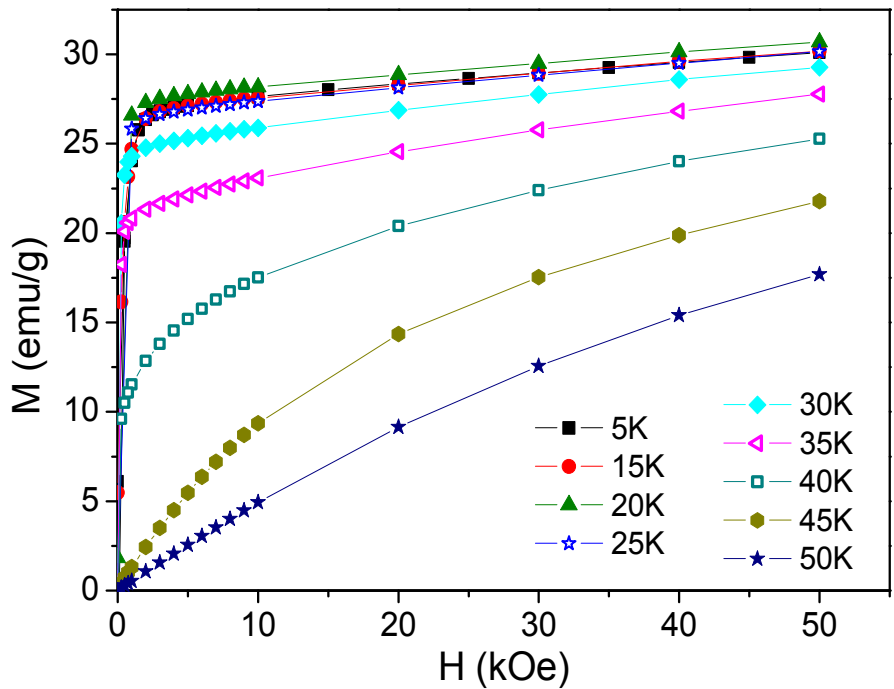

**Figure 2.** Magnetization versus applied field at different temperatures near and below $T_C$.

*3.2 Ferromagnetic resonance*

The FMR spectroscopy is a very sensitive technique to detect magnetic transitions as well as changes in the magnetic anisotropy of local-moment systems [25, 26], which are usually difficult to measure by other techniques, particularly in polycrystalline samples. Figure 3 (a) and (b) exhibit representative FMR spectra measured at different temperatures in the

T<$T_H$ and $T_H$<T $T_C$ ranges, respectively. For polycrystalline samples the resonance spectrum includes the contribution of the absorption lines of the crystallites oriented in all the possible space directions relative to the magnetic field. The main features observed in the temperature evolution of the spectra can be summarized as follows:

i) For temperatures above $T_C$ (i.e. in the paramagnetic phase) only one symmetric absorption line is observed, centered at an approximate constant value of $H_r$= 8619 Oe, corresponding to a spectroscopic splitting g-factor $g=(\hbar\omega)/(\mu_B H_r)$=1.991 (6), where $\hbar$ is the Planck's constant divided by $2\pi$ and $\mu_B$ is the Bohr's magneton.

ii) As the temperature decreases below $T_C$ the absorption line grows up, becomes asymmetric and the peak to peak linewidth, $\Delta H$, enhances. Furthermore, $H_r$ shifts to lower magnetic fields.

iii) Below T ~18 K more significant changes are detected: a secondary peak emerges and shifts to lower fields when the temperature diminishes.

These features could be explained by the presence of internal fields when the system goes through the magnetic transitions. In order to account for the temperature evolution of the spectrum, we introduce in the next section a phenomenological model that takes into account different terms that contribute to the free energy.

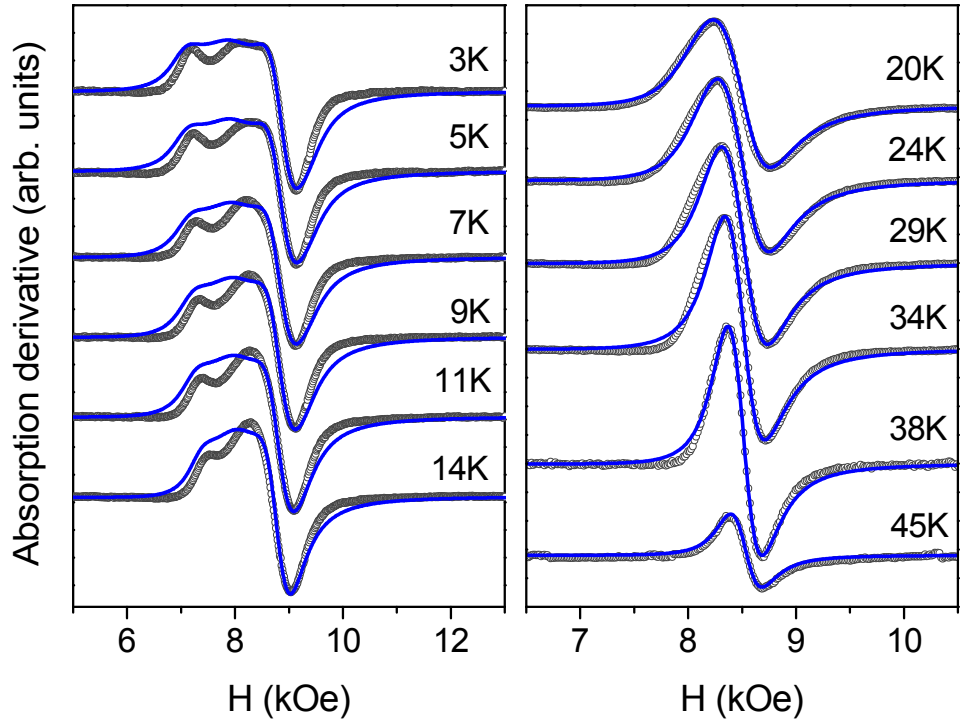

**Figure 3.** FMR absorption derivative spectra of the MnCr$_2$O$_4$ powder sample (open circles) measured at different temperatures: (a) T<$T_H$ and (b) $T_H$<T $T_C$. The straight lines correspond to the fittings with the phenomenological model.

*3.3 Evolution of the effective magnetic anisotropy: phenomenological model*

The ferromagnetic resonance condition is obtained from the magnetic free energy of the system following the Smit and Beljers formalism [27,28]. Equation (1) describes the different terms that contribute to the magnetic free energy $E$ of the MnCr$_2$O$_4$ system:

$$E = E_Z + E_{Kcub} + E_{Ku} \tag{1}$$

The first term of Eq. (1) corresponds to the Zeeman energy, described in Eq. (2), where $\vec{H} = H_0(\cos\varphi_H \sin\theta_H, \sin\varphi_H \sin\theta_H, \cos\theta_H)$ is the applied magnetic field vector, and

$\vec{M} = M_0(\cos\varphi\sin\theta, \sin\varphi\sin\theta, \cos\theta)$ is the magnetization vector in the laboratory coordinate system presented in Figure 4. The second term in Eq. (1) accounts for the second order cubic magnetocrystalline anisotropy, Eq. (3), characterized by the $K_1$ and $K_2$ anisotropy constants, where $r_1=\sin\theta\cos\varphi$, $r_2=\sin\theta\sin\varphi$ and $r_3=\cos\theta$. Finally, in order to consider the formation of the helicoidal phase, we have included a third term accounting for a uniaxial anisotropy characterized by the $K_u$ parameter, described by Eq. (4). This last term determines an easy axis in the $[1\bar{1}0]$ direction, which is the propagation direction reported for the helicoidal order [14].

$$E_Z = -\vec{M}\cdot\vec{H} = -M_0 H_0 [\cos\theta\cos\theta_H + \cos(\varphi-\varphi_H)\sin\theta\sin\theta_H] \qquad (2)$$

$$\begin{aligned}E_{Kcub} &= K_1(r_1^2 r_2^2 + r_2^2 r_3^2 + r_3^2 r_1^2) + K_2(r_1^2 r_2^2 r_3^2) = \\ &= K_1(\cos^2\theta\sin^2\theta + \cos^2\varphi\sin^2\varphi\sin^4\theta) + K_2\cos^2\varphi\sin^2\varphi\sin^4\theta\cos^2\theta\end{aligned} \qquad (3)$$

$$E_{Ku} = -\frac{K_u}{2}\sin^2\theta[1-\sin(2\varphi)] \qquad (4)$$

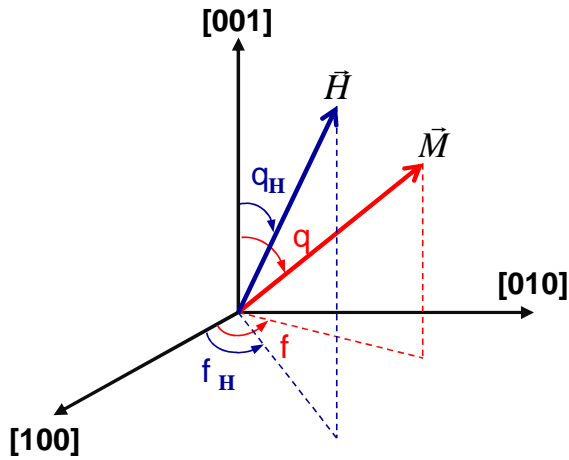

**Figure 4.** Schematic representation of the magnetization ($\vec{M}$) and magnetic field ($\vec{H}$) vectors and the angles involved in the description of the free energy.

Notice that the formalism applied to calculate the resonance mode of the ferrimagnetic $MnCr_2O_4$ spinel is essentially the same as the ferromagnetic resonance, in the sense that it is considered the precession of the spontaneous magnetization as a whole around their equilibrium orientation. Additional resonance modes, which depend explicitly on the magnetic sublattice structure, are not considered because they are located at frequencies much higher than the microwave. These modes involve the exchange interaction between the different magnetic sublattices which are usually above the infrared part of the spectrum. In fact, from the Mn-Mn, Cr-Mn and Cr-Cr exchange constants reported for the MnCr2O4 spinel the exchange resonance modes are above $\sim 5 \times 10^{11}$ s$^{-1}$, which is far from the microwave frequency range, and as a consequence these exchange modes could not be excited [23, 29]. Therefore, as it is usually implemented, we only include in the magnetic free energy the Zeeman interaction and the magnetic anisotropy terms. In the case of the cubic magnetocrystalline anisotropy, the easy direction of the magnetization depends on the signs and relative magnitude of $K_1$ and $K_2$. In the present case, we set $K_1<0$ and $9|K_1|/4<K_2<9|K_1|$ [30]. With this choice it results that the *easy*, *medium* and *hard* magnetization axis are parallel to the <110>, <111> and <100> directions of the crystal, respectively, for all temperatures. It is noteworthy that, if another relation between $K_1$ and $K_2$ is chosen (resulting in different medium and hard magnetization directions), this leads to qualitatively different spectra features, where secondary absorption peaks are localized in the g<2 higher field region. We also want to remark that the aforementioned choice of parameters is consistent with the magnetization easy axis direction reported from neutron diffraction and magnetization studies performed on single crystal samples [13,14,31] and differs from the results reported by [32] where different orientation of the easy magnetization direction was found.

Regarding $K_u$, this parameter takes into account the propagation direction of the helicoidal order [14,15], that breaks the cubic symmetry imposed by the crystalline structure. This kind of magnetic ordering is observed when several comparable exchange interactions are present and the description in terms of sublattices is interdicted. This is the case, for example, when the step of the spiral is not commensurate with the lattice parameter [33].

From the magnetic free energy, equations (1) to (4), the angular derivatives $(\partial^2 E/\partial \theta^2, \partial^2 E/\partial \varphi^2$ and $\partial^2 E/\partial \theta \partial \varphi)$, evaluated at the equilibrium angles for the magnetization for each orientation of the magnetic field, were calculated. The FMR resonance condition was obtained evaluating the Smit-Beljers equation [27,28]:

$$\left(\frac{\omega}{\gamma}\right)^2 = \frac{1}{M_0^2 \sin^2 \theta} \left[\frac{\partial^2 E}{\partial \theta^2}\frac{\partial^2 E}{\partial \varphi^2} - \left(\frac{\partial^2 E}{\partial \theta \partial \varphi}\right)^2\right]. \qquad (5)$$

Here ω is the angular frequency ω/2π ~ 24 GHz, γ is the gyromagnetic ratio and $M_0$ is the saturation magnetization value measured at ω/γ~ 8 kOe (Figure 2). As we measured a polycrystalline sample we assume that the absorption line corresponds to the sum of Lorentzian lineshape resonances with a *homogeneous* angular distribution of the anisotropies axes related to the magnetic field. For simplicity no angular variation of the resonance linewidth was considered. Furthermore, for temperatures near and below $T_C$ the lines present an additional asymmetry that could be attributed to a dispersive component

[34,35] as we are going to discuss later. Consequently, in this range we have also included in the simulated spectra a dispersive term, determining a lineshape of the form: (1-$\xi$) Absorption + $\xi$ Dispersion, where 0<$\xi$<1 [35-36]. We solved the Smit-Beljers equation (Eq. 5) in a self-consistent way, with g, $K_1$, $K_2$ and $K_u$ as adjusted parameters, and we have obtained a numerical simulation for the FMR resonance absorption at each temperature. The gyromagnetic factor obtained from the fittings in all the T $T_C$ range is g~2.05(2). The calculated spectra are presented in straight lines in figure 3, where good agreement between the spectral lines and the model is observed in all the studied temperature range. Notice that the calculated spectra reproduce well the general features of the lineshape, as the resonant field, the field positions of the satellite peaks and the linewidth, even for T<$T_H$ where the experimental peaks are broader than the fitting. The difference between the experimental and the calculated spectra could be attributed to the simplifications of our model, as we considered no angular variation of $\Delta H$ on the resonance lines that form the powder spectrum and also no distribution of anisotropy values were considered that could account for some degree of crystalline disorder. Nevertheless, this simple model allows us to extract quantitative information of the evolution of the system with temperature.

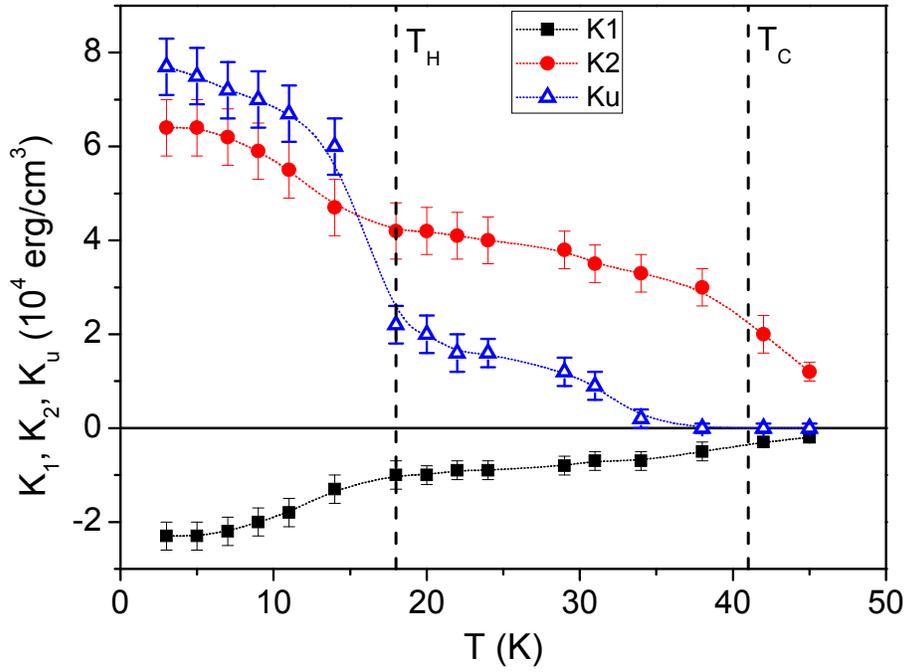

**Figure 5.** Temperature evolution of the anisotropy constants calculated from the phenomenological model. The lines are guides for the eye and the vertical dotted lines signal the ferrimagnetic order temperature $T_C$ and the helicoidal temperature $T_H$.

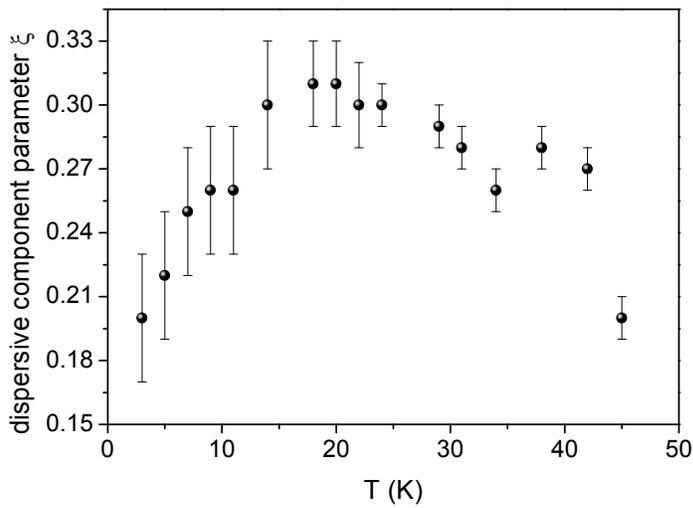

**Figure 6.** Temperature evolution of the parameter ξ that corresponds to the proportion of dispersive component included in the spectra simulations.

In figure 5 we have presented the temperature evolution of the magnetic anisotropy constants obtained from the simulation of FMR spectra. Below $T_C$, the magnitude of the cubic magnetocrystalline anisotropy constants, $K_1$ and $K_2$, starts to increase smoothly and shows a more important enhancement below $T_H$. These cubic anisotropy values are in the same order of magnitude than the values reported for similar oxide and chalcogenide spinel systems as $MnFe_2O_4$ ($K_1 \sim -3.3 \times 10^4$ erg/cm$^3$ [38, 39]) and $MnCr_2S_4$ ($K_1 \sim 4.2 \times 10^4$ erg/cm$^3$ and $K_2 \sim 1 \times 10^5$ erg/cm$^3$ [40]). On the other hand, $K_u$ remains equal to zero down to ~30K, where it starts to enhance smoothly. Finally, a jump in the $K_u$ value is observed below $T_H$, followed by a lineal increase up to the lowest measured temperature. The value of $K_u \sim 8 \times 10^4$ erg/cm$^3$ obtained at 4.2 K agrees with that reported in Ref. [41] of ~$2 \times 10^4$ erg/cm$^3$. The abrupt jump of $K_u$ is reflected in the low field satellite peak that clearly appears in the FMR spectra below $T_H$ [see figure 3(a)] which is related to the formation of the short-range helicoidal order. However, $K_u$ is non-zero above $T_H$, till T~30 K. This result could indicate that the short-range helicoidal order still coexists with the ferrimagnetic order above the helical transition temperature. Although no evidence of this fact was detected within statistical uncertainties from neutron diffraction experiments for this system, Tomiyasu and co-workers [14] observed that for the $CoCr_2O_4$ spinel, the spiral component retains the correlation well above $T_H$.

Furthermore, this result is consistent with the magnetization measurements where the high field lineal contribution is observed from low temperature up to T ~ 30 K. This complex stage is a consequence of exchange and superexchange competing interactions in this geometrically frustrated magnetic material. Although the relevant interaction is the direct Cr-Cr exchange interaction, in the $MnCr_2O_4$ the superexchange interactions between Cr-Mn and Mn-Mn present comparable magnitude [4]. Therefore the long-range helicoidal order cannot be stabilized, as is calculated for $AB_2O_4$ spinel when the A-A interaction is neglected [42,43]. These results suggest that, at low temperature, the long-range ferrimagnetism coexists with the spiral order where the transverse component of the Mn and Cr conical arrange is ordered. However, when the temperature increases the conical arrange preserves the longitudinal order up to $T_C$, while the transverse component is largely disordered above ~18 K.

Figure 6 exhibits the parameter $\xi$, that determines the proportion of dispersive component included in the fittings, as a function of temperature. Notice that this component is larger in the $T_H<T<T_C$ temperature range where the magnetization increases. It is well known that in a medium which has conductivity ($\sigma$) the wave is attenuated as it progresses through the sample [35-37]. Therefore, when the penetration depth of the microwave ($\delta$) is less than the thickness of the sample (d) the medium will not be homogeneous and the resonance line will be asymmetric. This effect is more important for materials with high magnetic permeability ($\mu$). For conductive materials this resonance correspond to a Dysonian line, which in the limit $0 < d/\delta < 2$ results a linear combination of absorption and dispersion Lorentzian lines [36]. In the case of insulator materials, as the $MnCr_2O_4$ [5, 15], the microwave penetration depth results: $\delta \approx \frac{2}{f}\sqrt{\frac{\nu}{\varepsilon}}$, where $\varepsilon$ corresponds to the dielectric constant [35]. Therefore, the microwave penetration depth decreases with the magnetic permeability, as a consequence the dispersive component in the magnetic resonance increases (see figure 1). This result is in agreement with the measured temperature dependence of the magnetization which is larger in the $T_H<T<T_C$ temperature range. However, the wave propagation in the medium depends on the interplay between the characteristic σ, ε and μ parameters. Recently it has been reported that multiferroic spinels present in the magnetic ordered phase important changes in the dielectric properties, besides the increases of the magnetization [44, 45]. Consequently, in order to study this topic in depth, careful measurements of the temperature evolution of these parameters in $MnCr_2O_4$ should be performed.

## 4. Conclusions

In summary, we have investigated the low temperature magnetic phases present in a polycrystalline sample of the $MnCr_2O_4$ spinel and quantified the temperature evolution of the magnetic anisotropy constants in a wide temperature range. In the magnetization versus temperature measurements we have observed anomalies consistent with the ferrimagnetic order at $T_C$~41K and the formation of the helicoidal spin arrangement at $T_H$~18 K. The electron spin resonance spectra exhibit important changes as a function of the temperature. This behavior could be explained through a phenomenological model considering the different terms that contribute to the magnetic free energy of the system. Below $T_C$ the FMR spectra can be fitted by a cubic magnetocrystalline anisotropy term with constants $K_1$ and $K_2$ that increase when the temperature diminishes. Near $T_H$, an additional magnetic anisotropy term should be included to account for the noticeable changes observed in the FMR spectra. This anisotropy, accounted by a $K_u$ parameter of the uniaxial anisotropy term, is associated to the breaking of the cubic anisotropy due to the formation of the helicoidal order that propagates in the [1$\bar{1}$0] direction. We remark that the structure observed in the FMR spectra can be explained taking into account the change of the magnetic symmetry of a single magnetic phase as a function of the temperature. The fact that $K_u$ is non-zero above $T_H$ could indicate that the short-range helicoidal order still coexists with the ferrimagnetic order above the helical transition. The fact that Ku is non-zero above $T_H$ and also the lineal increase of the high field magnetization up to 30 K, could indicate that the conical arrangement of the spins coexists with the ferrimagnetic order above $T_H$. When the temperature diminishes the transverse component of the Mn and Cr conical arrange orders and the spiral order stabilizes. Finally, we want to emphasize the sensitivity of the electron spin resonance spectroscopy to detect magnetic transitions and anisotropic interactions which enable us to obtain fundamental information that complements the magnetic measurements, and it allows us to calculate the characteristic parameters even in polycrystalline samples.

**Acknowledgements**

The authors thank Francisco Rivadulla for the samples and valuable suggestions and discussions. The work was supported by PIP 112-200801-0133 CONICET (Argentina) and C011 Universidad Nacional de Cuyo (Argentina).

**References**

[1] R. Moessner and J. T. Chalker 1998 Phys. Rev. B **58** 12049

[2] S. E. Dutton, Q. Huang, O. Tchernyshyov, C. L. Broholm and R. J. Cava 2011 Phys. Rev. B **83** 064407

[3] S.-H. Lee, C. Broholm, W. Ratcliff, G. Gasparovic, Q. Huang, T. H. Kim and S.-W. Cheong 2002 Nature **418** 856-858

[4] E. Winkler, S. Blanco Canosa, F. Rivadulla, M. A. López-Quintela, J. Rivas, A. Caneiro, M. T. Causa and M. Tovar 2009 Phys. Rev. B **80** 104418

[5] S. Blanco-Canosa, F. Rivadulla, V. Pardo, D. Baldomir, J.-S. Zhou, M. García-Hernández, M. A. López-Quintela, J. Rivas, and J. B. Goodenough 2007 Phys. Rev. Lett. **99** 187201

[6] H. Martinho, N. O. Moreno, J. A. Sanjurjo, C. Rettori, A. J. García-Adeva, D. L. Huber, S. B. Oseroff, W. Ratcliff II, S.-W. Cheong, P. G. Pagliuso, J. L. Sarrao, and G. B. Martins 2001 Phys. Rev. B **64** 024408

[7] H. Ueda, H. Mitamura, T. Goto, and Y. Ueda 2006 Phys. Rev. B **73** 094415

[8] S. -H, Lee, T. H. Kim, W. Ratcliff and S. -W. Cheong 2000 Phys. Rev. Lett. **84** 3718

[9] O. Tchernyshyov, R. Moessner and S. L. Sondhi 2002 Phys. Rev. Lett.**88** 067203

[10] M. Matsuda, H. Ueda, A. Kikkawa, Y. Tanaka, K. Katsumata, Y. Narumi, T. Inami, Y. Ueda, and S.-H. Lee 2007 Nature Phys. **3** 397-400

[11] Y. Tanaka, Y. Narumi, N. Terada, K. Katsumata, H. Ueda, U. Staub, K. Kindo, T. Kukui, T. Yamamoto, R. Kammuri, M. Hagiwara, A. Kikkawa, Y. Ueda, H. Toyokawa, T. Ishikawa, and H. Kitamura 2007 J. Phys. Soc. Jpn. **76** 043708

[12] S. Bordács, D. Varjas, I. Kézsmárki, G. Mihály, L. Baldassarre, A. Abouelsayed,

C. A. Kuntscher, K. Ohgushi, and Y. Tokura 2009 Phys. Rev. Lett.**103** 077205

[13] J. M. Hastings and L. M. Corliss 1962 Phys. Rev. **126** 556-565

[14] K. Tomiyasu, J. Fukunaga and H. Suzuki 2004 Phys. Rev. B **70** 214434

[15] N. Menyuk, K. Dwight and A. Wold 1964 J. Phys. (Paris) **25** 528-536

[16] D. I. Khomskii 2006 J. Magn. Magn. Mater. **306**1-8

[17] T. Arima, Y.Yamasaki, T. Goto, S. Iguchi, K. Ohgushi, S. Miyasaka and Y. Tokura 2007 J. Phus. Soc. Japan **76** 023602

[18] S. -W. Cheong and M. Mostolov 2007 Nature Matt. **6** 13-20

[19] N. Mufti, G. R. Blake and T. T. M. Palstra 2009 J. Magn. Magn. Mater. **321** 1767-1769

[20] N. Mufti, A. A. Nugroho, G. R. Blake and T. T. M. Palstra 2010 J. Phys.: Condens. Matter **22** 075902

[21] Y.Huang, Z. Qu and Y.Zhang 2011 J. Magn. Magn. Mater. 323970

[22] R. N. Bhowmik, R.Ranganathan and R.Nagarajan 2006 Phys. Rev. B **73** 144413

[23] A. H. Morrish, The Physical Principles of Magnetism (IEEEPress, NY, 2001).

[24] I. S. Jacobs 1960 J. Phys. Chem. Solids **15** 54

[25] G. Alejandro, J. Milano, L. B. Steren, J. E. Gayone, M. Eddrief and V. H. Etgens 2012 Physica B **407** 3161-3164

[26] E. Winkler, M. T. Causa and C. A. Ramos 2007Physica B **398** 434-437

[27] J. Smit and H.G. Beljers 1955 Philips Res. Rep. **10** 113

[28] C. Vittoria, *Microwave Properties of Magnetic Films* (World Scientific Pub. Co. Inc., Singapore, 1994).

[29] J. Milano, L. B. Steren and M. Grimsditch 2044 Phys. Rev. Lett. **93** 077601

[30] B. D. Cullity and C. D. Graham, *Introduction to Magnetic Materials* (John Wiley & Sons, Inc., 2nd Edition, New Jersey, 2009).

[31] T. Tsushima, Y. Kino, and S. Funahashi 1968 J. Appl. Phys. **39** 626

[32] S. Funahashi, KiitiSiratori and Y. Tomono 1970 J. Phys. Soc. Japan **29** 1179-1193

[33] J. S. Smart, "*Effectivefieldtheories of magnetism*" (W. B. Saunders Company, USA, 1966)

[34] R. Jarrier, X. Marti, J. Herrero-Albillos, P. Ferrer, R. Haumont, P. Gemeiner, G. Geneste, P. Berthet, T. Schülli, P. Cevc, R. Blinc, S. S. Wong, Tae-Jin Park, M. Alexe, M. A. Carpenter, J. F. Scott, G. Catalan and B. Dkhil 2012 Phys. Rev. B **85** 184104

[35] E. C. Jordan, "*Electromagnetic waves and radiating systems*" (Prentice-Hall, Inc., USA, 1964).

[36] L. Walmsley 1996 J.Magn.Reson. A **122** 209

[37] C. P. Poole Jr., *Electron Spin Resonance: A Comprehensive Treatise on Experimental Techniques* (Courier Dover Publications, 1996).

[38] R. S. Weisz 1954 Phys. Rev. **96** 800-801

[39] X. Zuo, A. Yang, S. Yoon, J.Christodoulides, V. G. Harris, and C. Vittoria 2005 Appl. Phys. Lett. **87** 152505

[40] V. Tsurkan, M. Mücksch, V. Fritsch, J. Hemberger, M. Klemm, S. Klimm, S. Körner, H.-A. Krug von Nidda, D. Samusi, E.-W.Scheidt, A. Loidl, S. Horn and R. Tidecks 2003 Phys. Rev. B **68** 134434

[41] S. Krupi ka, Z. Jirák, P. Novák, F. Zounová and V. Roskovec 1980 Acta Phys. Slov. 30 251.

[42] D. H. Lyons, T. A. Kaplan, K. Dwight, and N. Menyuk 1962 Phys. Rev. **126** 540

[43] J. B. Goodenough, *Magnetism and Chemical Bond* (Wiley, New York, 1963).

[44] Y. Yamasaki, S. Miyasaka, Y. Kaneko, J.-P.He, T. Arima, and Y. Tokura 2006 Phys Rev. Lett.**96** 207204

[45] E. D. Mun, Gia-Wei Chern, V. Pardo, F. Rivadulla, R. Sinclair, H. D. Zhou, V. S. Zapf, and C. D. Batista 2014 Phys Rev. Lett. **112** 017207